# Z-AXIS OPTOMECHANICAL ACCELEROMETER

*D.N. Hutchison and S.A. Bhave*
OxideMEMS Laboratory, Cornell University, Ithaca NY, USA

## ABSTRACT
We demonstrate a z-axis accelerometer which uses waveguided light to sense proof mass displacement. The accelerometer consists of two stacked rings (one fixed and one suspended above it) forming an optical ring resonator. As the upper ring moves due to z-axis acceleration, the effective refractive index changes, changing the optical path length and therefore the resonant frequency of the optical mode. The optical transmission changes with acceleration when the laser is biased on the side of the optical resonance. This silicon nitride "Cavity-enhanced OptoMechanical Accelerometer" (COMA) has a sensitivity of 22 percent-per-g optical modulation for our highest optical quality factor ($Q_o$) devices.

## INTRODUCTION
### Motivation
Optics has been shown to yield extremely sensitive readout of mechanical motion, achieving sub-fm/Hz$^{1/2}$ sensitivity at room temperature in a variety of previous on-chip devices [1-2], and can be more sensitive to mechanical displacement than electrostatics for a given transduction area. In particular, the optical resonance frequency of two vertically-stacked ring cavities has been shown to be exquisitely sensitive to the gap between them [3-4]. In previous stacked-ring devices, both rings were released — those devices were designed for optical wavelength routing and are not sensitive to acceleration. By anchoring one ring to the substrate and suspending the other ring above it, we exploit this gap-change sensitivity for inertial measurement.

### The COMA
The device consists of two stacked silicon nitride rings as shown in Fig. 1. The upper ring is anchored in the center but its rim is free to move, while the bottom ring is fixed to the substrate. Light is evanescently coupled into the rings and circulates around their outermost rims. The rings are close enough that there is optical coupling between them and they support gap-sensitive optical "supermodes" (Fig. 1 (C)). Only optical wavelengths that fit an integer number of times around the circumference are supported in the stacked-ring resonator, and other wavelengths pass by. Acceleration causes the ring-ring gap to change, changing the optical resonance frequency, which is measured as optical intensity variation at the output.

For optomechanical systems such as this, an important figure of merit is the optomechanical coupling constant $g_{om} = \frac{d\omega_o}{dz}$ [5]. This measures how much the optical resonance frequency $\omega_o$ changes with mechanical deflection $z$. In the Modeling section, we calculate $g_{om}$ for our device and find it to be similar to similar stacked-ring devices reported previously.

Although the proof mass (the upper ring) is small (total mass of only 8.7 ng), an arbitrarily large mass can be attached to the upper ring by, for instance, a back-side through-wafer etch of the underlying substrate. The experiments reported here simply demonstrate the sensing mechanism and show that it is effective despite the small proof mass.

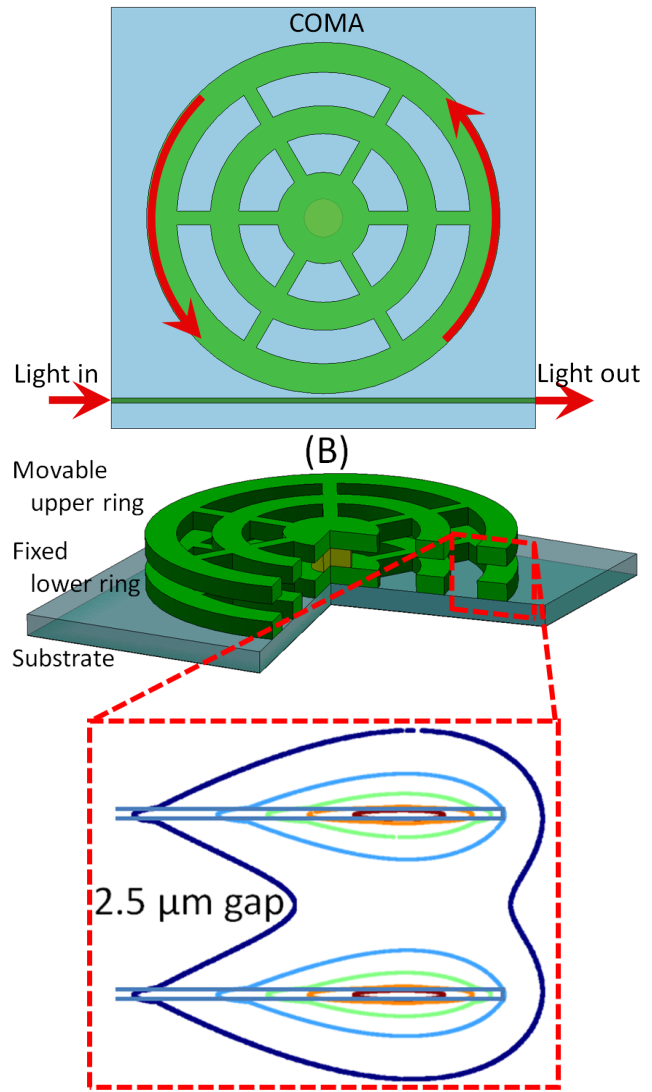

*Figure 1: (A) Top, and (B) perspective schematic view of a COMA (thickness and gap exaggerated for clarity). For devices presented in this paper, radius is 100 µm, ring and spoke widths are 8 µm, thickness is 220 nm, and gap at rim after release is 2.5 µm (C) Contour plot of the electric field intensity of the simulated fundamental optical supermode. Contour scale is linear but in arbitrary units, and thickness and gap are drawn to scale.*

### Relation to Prior Work
Optomechanical light modulation has been achieved in a variety of devices. In one embodiment, light passing through a photonic crystal waveguide is modulated using a nearby cantilever [6-7]. In another incarnation, an optical resonance shift due to mechanical actuation was reported [3-4] for stacked ring resonators similar to our devices. In contrast to the former references, these latter schemes are cavity-enhanced—they depend on an optical resonance and the optical quality factor of the cavity. Although these cavity-enhanced schemes give excellent displacement sensitivity, to our knowledge they have not previously been used for inertial measurement.

A different kind of optomechanical accelerometer was proposed [8] but to our knowledge has not been demonstrated. In that device, acceleration changes the distance between a waveguide and an optical disk resonator which changes the *coupling coefficient*, changing the total optical quality factor and thereby the transmitted light intensity at a given wavelength. In contrast, our device relies on shifting the *resonant frequency* of the cavity itself, preserving the quality factor and only shifting it in frequency.

An on-chip optomechanical accelerometer was demonstrated [9] but it uses free space laser light and therefore it is not clear how laser, device, and detector can be integrated on a single chip. Since our device uses waveguided light, our device could be entirely integrated on a single chip with lasers (such as VCSELs with optical coupling gratings [10]) and detectors (such as [11]).

In another previous device, an optical grating was fabricated on the proof mass of a traditional capacitive MEMS accelerometer, and a free-space laser and off-chip detector used, forming a nano-g accelerometer with displacement sensitivities as low as 12 fm/√Hz at 1 kHz [12-13]. Although this proves the extreme sensitivity of optics for sensing displacement, it too relies on free-space lasers and it is not clear it can be implemented on a single chip.

## MODELING
### Static Deformation and Mechanical Frequency

Finite-element simulations show that under 1 g of acceleration, the 100 μm radius, 220 nm thick silicon nitride upper ring will deform by around $D = 0.45$ nm/g at the rim (Fig. 2).

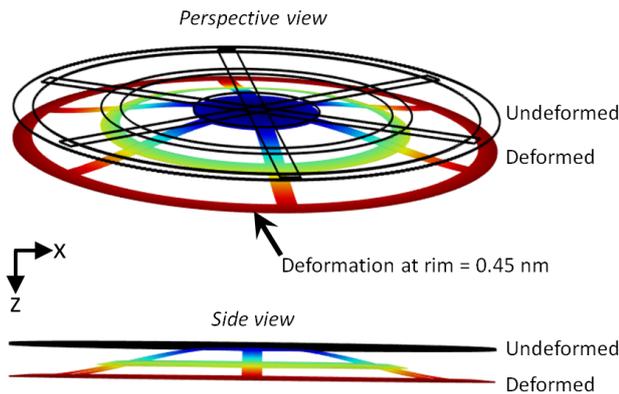

*Figure 2: COMSOL simulation showing expected deformation of the upper ring due to gravity. Only the upper (released) ring is shown, in both undeformed (black wireframe) and exaggerated deformed (colored surface) state.*

The mechanical resonant frequency of this structure is also found using finite-element simulations to be $\omega_m = 65$ kHz. (In this paper we use subscript "*m*" to denote mechanical and subscript "*o*" to denote optical quantities.)

### Optical Mode

Next we simulate the optical mode to calculate how the optical resonant frequency will be affected by deformation of the upper ring. We model the optical mode of the stacked rings using finite element software (Fig. 1 (C)) to extract the optical eigenfrequencies $\omega_o$ versus ring-ring gap $z$ for the fundamental optical mode. The resonant frequency and its derivative, $g_{om} = \frac{d\omega_o}{dz}$, is plotted in Fig. 3. For the gap in Fig. 1, we find $\frac{g_{om}}{2\pi} \approx 4$ GHz/nm.

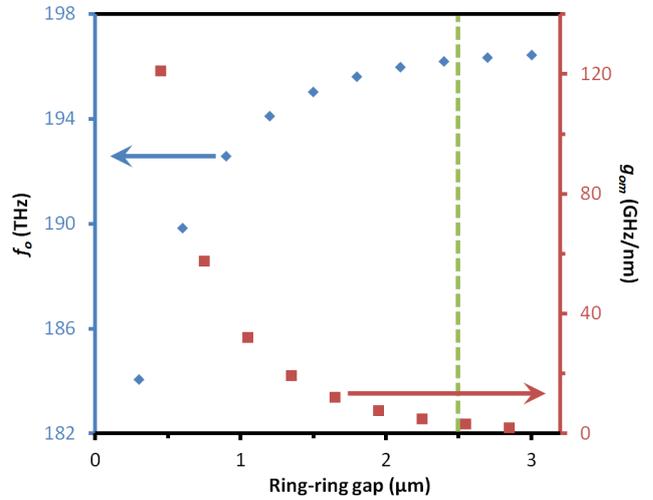

*Figure 3: Simulated optical eigenfrequency (blue diamonds) and optomechanical coupling strength $g_{om}$ (red squares) [5] for stacked rings separated by different gaps. The green dashed line indicates the measured gap for the device in Fig. 1.*

### Sensitivity to Acceleration

Knowing $g_{om}$, we calculate the expected shift in optical resonance wavelength $\lambda_o$ due to small movements $\Delta z$ of the upper ring. This is given by

$$\Delta\lambda_o \approx c\Delta f/f^2 = c\left(\frac{g_{om}}{2\pi}\Delta z\right) / \left(\frac{c^2}{\lambda_o^2}\right) = \frac{g_{om}\Delta z \lambda_o^2}{2\pi c} \quad (1)$$

From Fig. 2 we see that an acceleration change of 2 g (+g to –g) will cause a deflection of $\Delta z = 0.45$ nm × 2 = 0.90 nm at the rim. Using $g_{om} = 4$ GHz/nm and $\lambda_o = 1532.5$ nm, we find that the optical resonant wavelength will increase by 4.5 pm under a –g to +g acceleration change (achieved by flipping the chip). Although this is a small expected shift in resonant frequency, it is comparable to the FWHM of the first measured optical resonance we report in the results section, 23 pm.

Next we find the sensitivity of the device when the laser is placed at the point of steepest slope of the optical resonance. The transmitted power as a function of frequency $\omega$ for an optical ring resonator is [14]

$$P(\omega) = P_0 \left[ \frac{4(\omega-\omega_o)^2 + \omega_o^2 \left(\frac{1}{Q_{int}} - \frac{1}{Q_{coup}}\right)^2}{4(\omega-\omega_o)^2 + \omega_o^2 \left(\frac{1}{Q_{int}} + \frac{1}{Q_{coup}}\right)^2} \right] \quad (2)$$

where $P_0$ is the input power, and $Q_{int}$ and $Q_{coup}$ are the intrinsic optical quality factor and the coupling contribution, respectively. For critical coupling, $Q_{int} = Q_{coup}$, so the total optical quality factor $Q_o$ is

$$\frac{1}{Q_o} = \frac{1}{Q_{int}} + \frac{1}{Q_{coup}} = \frac{\Delta\omega_{FWHM}}{\omega_o} = \frac{\Delta\lambda_{FWHM}}{\lambda_o}. \quad (3)$$

The slope of the Lorentzian is greatest at an offset of $\frac{1}{\sqrt{3}}$ times HWHM:

$$\left.\frac{dP(\omega)}{d\omega}\right|_{\omega=\omega_o \pm \frac{1}{\sqrt{3}}\frac{\omega_o}{2Q_o}} = \frac{196\sqrt{3}}{2401} \frac{Q_o}{\omega_o} P_0 \equiv \xi \frac{Q_o}{\omega_o} P_0. \quad (4)$$

Rearranging the differentials and including acceleration we get

$$\frac{dP}{da} = \xi \frac{Q_o}{\omega_o} P_0 \frac{d\omega}{dz} \frac{dz}{da} \quad (5)$$

$$\frac{d(P/P_0)}{da} = \xi \frac{Q_o}{\omega_o} g_{om} D \quad (6)$$

for acceleration $a$ and vertical mechanical displacement $z$ of the upper ring.

## FABRICATION

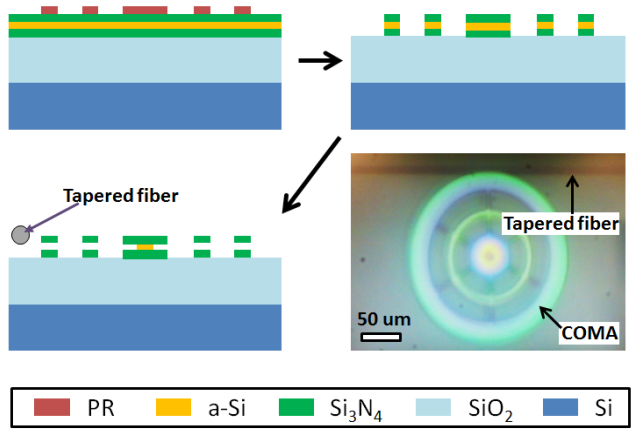

Figure 4: Process flow and optical microscope image of a released device. A spoked wheel shape is etched through a nitride : a-Si : nitride stack by RIE. (The underlying 4 μm oxide provides optical isolation.) The a-Si layer is then etched in XeF₂. The upwards curling of the upper layer after release (not shown in the schematic) prevents the upper layer from sticking to the lower nitride layer during XeF₂ release. The device is interrogated with a tapered optical fiber.

The COMA is made by growing 4 μm of oxide on Si, then depositing 220, 180, 220 nm of Si₃N₄, amorphous Si, Si₃N₄ respectively using LPCVD (Fig. 4). A 100 micron radius spoked wheel shape is etched through the three top layers with RIE (CHF₃/O₂ chemistry) and the a-Si sacrificial layer is removed by XeF2 to release the upper ring. After release the spokes curl upwards due to stress, resulting in a final ring-ring gap which is larger than the original 180 nm—for the device in Fig. 4 the gap was about 2.5 um. As shown in Fig. 3, smaller gaps are preferable; however even with this large gap we show good acceleration sensitivity.

## RESULTS
### Testing setup

The testing setup is shown in Fig. 5. For the swept-wavelength measurements, the laser (New Focus Velocity 6328) wavelength is swept continuously up and down, and the light is passed through a polarization controller and then split. One part is evanescently coupled from a tapered optical fiber (previously manufactured in-situ, similar to [15]) into the COMA and the transmitted light monitored on detector D1 (ILX Lightwave for Fig. 6, New Focus 1811 for Fig. 7). The other part passes through a fixed-length Fabry-Pérot cavity (Thorlabs SA210-12B) which provides fixed reference wavelengths to remove sweep-to-sweep drift of the laser (done during post-processing of data) and is monitored on detector D2 (Thorlabs SA210-12B) and an oscilloscope. For the mechanical spectrum measurements, the wavelength is fixed and a spectrum analyzer detects the optical modulation at D1. All light is contained in fiber optics except for the Fabry-Pérot cavity arm.

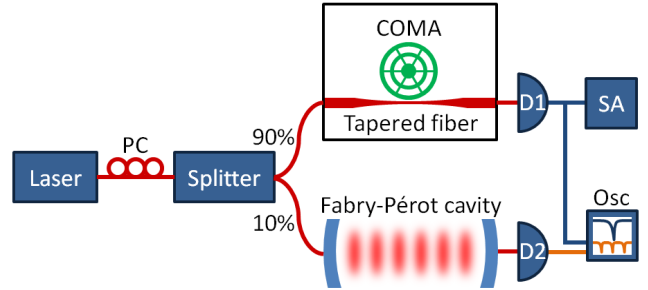

Figure 5: Testing setup. For the swept-wavelength measurements, the wavelength of the laser light is swept and the light is passed through a polarization controller (PC) and then split. Detectors D1 and D2 record the optical transmission spectra of the COMA and a reference cavity, and are monitored on an oscilloscope (Osc). For the mechanical spectrum measurements, the wavelength is fixed and D1 is monitored on a spectrum analyzer (SA).

### Static ± g Measurements

We measure the optical transmission of the device as a function of wavelength (Fig. 6). This optical mode has quality factor $Q_o = 66,000$. The +g (chip face up) resonance is at longer wavelengths than the –g (chip face down) measurement, because a smaller gap leads to a lower optical eigenfrequency as shown in Fig. 3, which corresponds to a longer resonant wavelength.

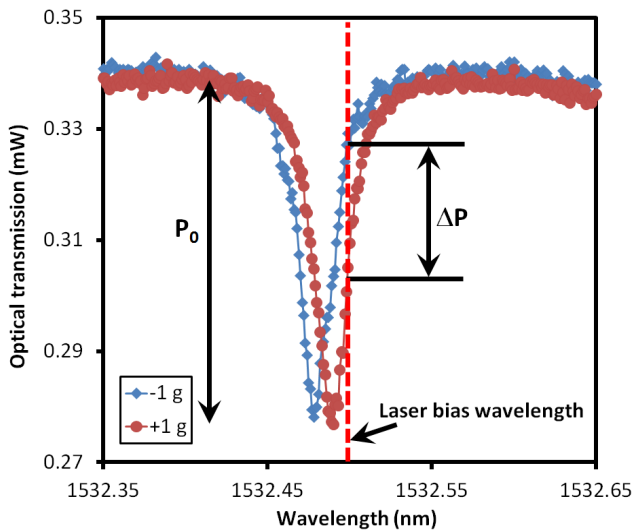

*Figure 6: Optical transmission spectrum for +g (red circles) and -g (blue diamonds) configurations. The half-maximum of one resonance lies at approximately the minimum of the other resonance, demonstrating large optical transmission change with only a few g's of force. The optical quality factor is $Q_o = 66,000$, FWHM is 23 pm, and the peak shifts by 10 pm under 2 g acceleration.*

Using Eq. 6 and $g_{om}$ and $D$ from finite element analysis, the theoretical wavelength shift is 4.5 pm and the sensitivity to acceleration for the mode shown here is 14 percent-per-g. This sensitivity value means that with the laser on the optical resonance at the point of steepest slope, for small accelerations, the light changes by 14% of its maximum to its minimum value. The observed shift is larger, at around 10 pm. The discrepancy in the measured resonant frequency shift probably comes from initial gap measurement uncertainty and a spring stiffening effect caused by up-curl after release of the upper ring due to stress.

The spectra shown here do not drop to zero transmission exactly at the resonant wavelength because (1) there is some mismatch in the effective refractive index between the tapered fiber and the COMA, and (2) we operate exclusively in the undercoupled regime in order to reduce the effect of coupling on resonant frequency. In this regime we observe the resonant frequency to only vary with coupling by at most around 0.002 nm. Beyond critical coupling the resonant frequency varies a lot more (nearly 1 nm) with different coupling strengths. It is however possible to match effective indices and operate at critical coupling using an integrated waveguide. At critical coupling, the transmission can theoretically drop to exactly zero [14], increasing $P_0$ and thus increasing the total amount of optical variation with acceleration. In that case, $P_0$ would represent the total off-resonant transmitted power rather than the difference between off-resonance to on-resonance transmitted power.

**Arbitrary Angle and Repeatability Measurements**

To test the resonant shifts for repeatability, we put the device on a tilt stage and tested at 0° (face up) and 65° tilt repeatedly. Each time we couple to the device it takes several minutes to reconfigure the setup and realign the chip and tapered fiber. Even after this time and even for the different arbitrary coupling strengths we chose in each measurement, the resonance shows a definite shift for 0° to 65° tilt (Fig. 7). While the resonant frequency does vary slightly even among measurements at the same tilt, these discrepancies seem to be due to differences in the coupling position of the tapered fiber, which may be mitigated in the future by using an integrated waveguide.

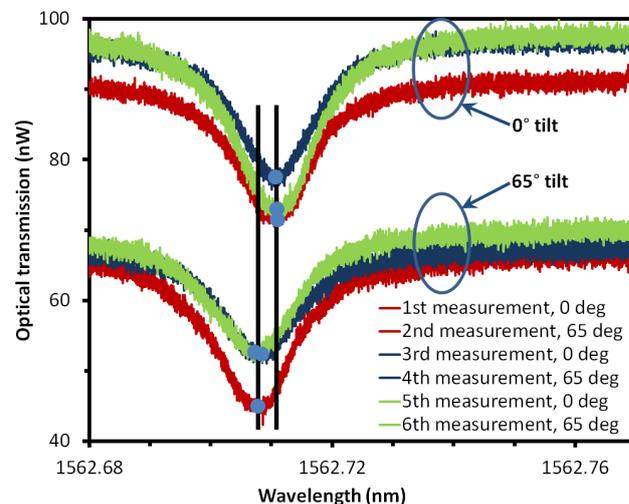

*Figure 7: Arbitrary angle repeatability measurements. The COMA was measured at 0°, then 65°, then 0° tilt, etc, repeatedly. Each time the taper would have been in a slightly different position and so a different amount of light was coupled in. Nevertheless a definite resonance shift is apparent, showing that the measured resonance shift is not due to differences in coupling strength from experiment to experiment. Vertical black lines and blue dots are overlaid to indicate the frequency shift.*

The measurements reported in this section were for a different device, and showed even higher optical Q of around 104,000. This increases the sensitivity from 14 to 22 percent-per-g compared with the optical resonance in Fig. 6, however here we also used a faster photodetector that picked up higher frequency noise—noise that was averaged out by the lower bandwidth power meter used in the Fig. 6 measurements.

**Mechanical Spectrum**

We measured the power spectral density of the device in air and vacuum to get an estimate of the bandwidth and noise floor of the COMA. However, the mechanical spectrum was not detectable in air due to squeeze-film damping (for a similar device mechanical quality factor in air was found to be only $Q_m \approx 2$ [16]). The mechanical spectrum of a COMA with four spokes (rather than six) was measured in vacuum (40 µTorr) and shown in Fig. 8. Due to limitations with fine-positioning of the taper in our vacuum chamber, in this measurement the taper is actually touching the device, and therefore the mechanical signal is small because the optical mode is degraded, and broad ($Q_m = 160$) because the taper damps the ring's mechanical motion. The resonant frequency is higher (40 kHz) than predicted by finite element analysis (24 kHz), due to the taper touching the device, and spring-stiffening effect caused by the up-curl discussed earlier. Non-contact measurements are ongoing to accurately estimate the noise floor.

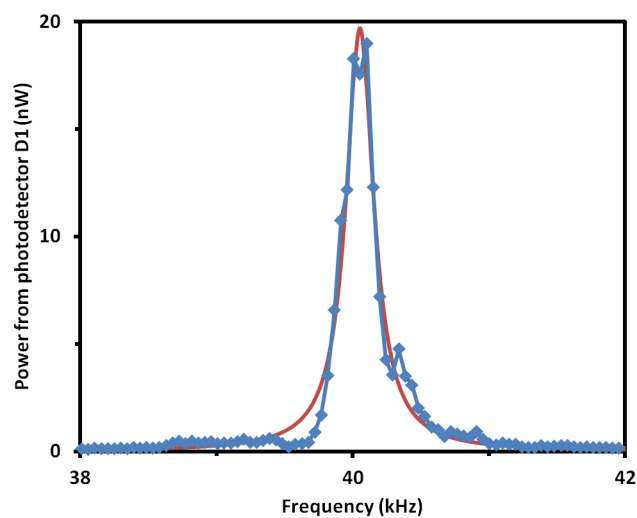

*Figure 8: COMA mechanical frequency in vacuum (blue diamonds) and Lorentzian fit (red curve), showing $Q_m$ = 160. The taper was touching the upper ring in these vacuum measurements, damping the mechanical motion.*

## CONCLUSION

We have demonstrated a cavity-enhanced optomechanical accelerometer using waveguided light that senses acceleration as a gap change between two stacked rings, changing the rings' optical resonant frequency. The optomechanical coupling constant is calculated and measured. The optical resonance shift, and the sensitivity of this inertial measurement scheme (which depends on the stiffness, mass, gap, and optical quality factor) are also calculated, and the sensitivity is found to be 22 percent-per-g for our highest-Q devices.

We observe resonance shifts of the correct magnitude and direction for static acceleration measurements, and find these shifts to be repeatable from measurement to measurement, even with slight differences in coupling of the input light. We also measure the mechanical spectrum in vacuum although in this measurement the coupling taper is touching the rim of the upper ring, which damps its mechanical motion.

Since we use waveguided rather than free-space light, the COMA has the potential to be integrated with other components on a single chip. The inertial measurement scheme proposed here could serve as the basis for an extremely sensitive, non-capacitive, all-dielectric accelerometer.

## ACKNOWLEDGEMENTS

We acknowledge the DARPA ORCHID program and the Intel Academic Research Office for funding. Devices were made in the Cornell Nanofabrication Facility.

Thank you to Siddharth Tallur, Laura Fegley, and Suresh Sridaran for stimulating discussions and for reading this paper through and offering advice.

## CONTACT

*D.N. Hutchison, tel: +1-213-537-6266; dnh37@cornell.edu